\documentclass[aps,pre,twocolumn,showpacs,superscriptaddress,amsmath,amssymb,longbibliography]{revtex4-2}

\usepackage[normalem]{ulem}
\usepackage{graphicx}
\usepackage{color,soul}
\usepackage{float}

\usepackage{mathtools}
\usepackage{hyperref}

\usepackage{xcolor}

\makeatletter 
\def\p@figure{\color{blue}} 
\def\p@equation{\color{blue}}
\def\p@table{\color{blue}}
\def\p@bibliography{\color{blue}} 
\makeatother

\graphicspath{ {./Figures/} }

\begin{document}

\title{The role of disorder in the motion of chiral swimmers in the presence of obstacles}

 \author{Danne M. van Roon}
  \affiliation{Centro de F\'isica Te\'orica e Computacional, Faculdade de Ci\^encias, Universidade de Lisboa, 1749-016 Lisboa, Portugal }
  \affiliation{Departamento de F\'isica, Faculdade de Ci\^encias, Universidade de Lisboa, 1749-016 Lisboa, Portugal}
  \author{Giorgio Volpe}
  \affiliation{Department of Chemistry, University College London, 20 Gordon Street, London WC1H 0AJ, UK. }
  \author{Margarida M. Telo da Gama}
  \affiliation{Centro de F\'isica Te\'orica e Computacional, Faculdade de Ci\^encias, Universidade de Lisboa, 1749-016 Lisboa, Portugal }
  \affiliation{Departamento de F\'isica, Faculdade de Ci\^encias, Universidade de Lisboa, 1749-016 Lisboa, Portugal}
  \author{Nuno A. M. Ara\'ujo}
  \affiliation{Centro de F\'isica Te\'orica e Computacional, Faculdade de Ci\^encias, Universidade de Lisboa, 1749-016 Lisboa, Portugal }
  \affiliation{Departamento de F\'isica, Faculdade de Ci\^encias, Universidade de Lisboa, 1749-016 Lisboa, Portugal}

\begin{abstract}
The presence of obstacles is intuitively expected to hinder the diffusive transport of micro-swimmers. However, for chiral micro-swimmers, a low density of obstacles near a surface can enhance their diffusive behavior, due to the rectification of the chiral motion by the obstacles. Here, we study numerically the role that disorder plays in determining the transport dynamics of chiral micro-swimmers on surfaces with obstacles. We consider different densities of regularly spaced obstacles and distinct types of disorder: noise in the dynamics of the micro-swimmer, quenched noise in the positions of the obstacles as well as obstacle size polydispersity. We show that, depending on the type and strength of the disorder, the presence of obstacles can either enhance or hinder transport, and discuss implications for the control of active transport in disordered media.
\end{abstract}

\maketitle
\section{Introduction}
\begin{figure*}[ht!]
     \centering
     \includegraphics[width=17.1 cm]{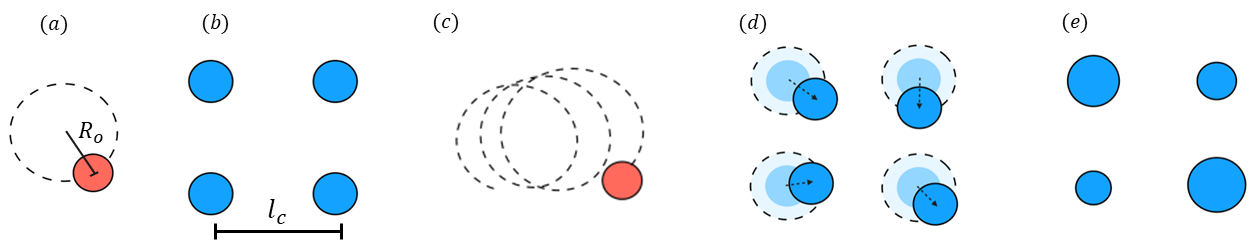}
        \caption{(a) Schematic depiction of an `ideal chiral swimmer' and its trajectory with orbit radius $R_{\textrm{o}}$. (b) Schematic depiction of four obstacles in a square lattice with lattice periodicity $l_{\textrm{c}}$. (c) Schematic depiction of the trajectory of a chiral swimmer with low dynamic noise. (d) Lattice with quenched noise in the form of randomly perturbed obstacle positions. The arrows denote the change in position of the obstacles (dark blue circles) due to the positional quenched noise, relative to their original lattice position marked by the immediately lighter shade of blue. The lightest shade of blue marks the area from which a new position was chosen randomly. (e) Lattice with quenched noise in the form of obstacle polydispersity.}
        \label{fig_1}
\end{figure*}
In the last two decades, active matter has become an increasingly important focus of research. On small length scales, active matter describes micron-sized entities such as motile cells and artificial micro-swimmers including Janus particles and active droplets \cite{rev}. Typically a micro-swimmer is represented as an active Brownian particle, or a particle undergoing run-and-tumble dynamics \cite{activematter}. Due to hydrodynamic interactions, several flagellated swimmers, such as \textit{E. coli} trace circular trajectories when swimming near a substrate \cite{whychiral, whychiral2, whychiral3}. Significant progress has been made in capturing the behavior of chiral micro-swimmers in homogeneous environments. Present tools include agent based models and various stochastic descriptions \cite {msd3, msd4, msd6, msd5}.

Biological and man-made micro-swimmers, however, rarely move in homogeneous environments, but rather encounter heterogeneity, such as domain walls, pores and obstacles \cite{rev}. 
To address this, recent research efforts have been directed at understanding the underlying (bio)physical mechanisms of micro-swimmers in heterogeneous environments \cite{steering}. From a fundamental perspective, this is interesting to understand and optimise search strategies in realistic environments \cite{search}. From a practical perspective, such an understanding is crucial to explain and control biomedically relevant processes, such as bio-film formation. In addition, there remains signiﬁcant yet unexploited potential to enable novel nanotechnological applications, including smart self-propelled cargo carriers with uses in drug-delivery in tissue or contamination removal in porous soil, among others \cite{application1, application2, application3, application4, environment1}.  

A first approach to control the dynamics of micro-swimmers in heterogeneous environments relies on designing the topography of the environment by carefully placing obstacles or structures on a substrate. Another approach would be to design a micro-swimmer that is able to navigate a complex environment in a controlled manner\cite{controlldr3}. This requires control over the reorientation of the swimmer, which has to date been achieved only in a limited number of cases \cite{controlldr1, controlldr2}.

Experimental observations show that the topography of the environment can strongly influence the dynamics of micro-swimmers, and in non-intuitive ways. Recent evidence indicates that the presence of porous micro-structures generally hinders the diffusive transport of micro-swimmers \cite{experiments1, lorentzgas}. Interestingly, for chiral micro-swimmers, contrasting phenomenology has also been observed. For example, a significantly enhanced propagation on surfaces, due to randomly placed obstacles, has been reported in theoretical studies in \cite{franosh1, franosh2} and experimentally for \textit{E. coli} in \cite{forward}. Furthermore, experiments tracking \textit{E. coli} navigating a colloidal crystal found that the colloids rectified the trajectory of bacteria, resulting in enhanced transport \cite{lattice, lattice2, lattice3, 3d}.

A common feature of natural and man-made heterogeneous environments at the micro-scale is the presence of disorder or noise in the surface topography. Disorder can occur in the form of spatial positioning and size of obstacles or pores. Also the motion of micro-swimmers contains disorder in the reorientation, typically in the form of flagellar noise or rotational Brownian diffusion\cite{controlldr1, controlldr2, noiseflagella, noiseflagella2}.

In this article, we examine numerically the influence of disorder (from here on referred to as `noise') on the dynamics of chiral micro-swimmers exploring a periodic arrangement of obstacles, where we introduce noise in a controlled manner. We first consider a chiral swimmer that does not experience any rotational Brownian diffusion in a periodic array of obstacles. Distinct types of noise are then introduced independently: `dynamic noise' determining the reorientation of the swimmer and two types of `quenched noise', the first being disorder in the positions of the obstacles, and the second being polydispersity in the sizes of the obstacles. 

To quantify the effect of the different types of noise on the transport properties of the swimmer, we computed its effective diffusion coefficient (diffusivity) for a wide range of obstacle area fractions (densities) and noise strengths. We find that the diffusivity of a chiral swimmer is strongly non-monotonic for increasing density of obstacles. At lower densities, the transport of a chiral swimmer is rectified due to the interaction of the swimmer with the obstacles, affecting the swimming orbits, and allowing the swimmer to explore space. At higher densities, the swimmer is unable to perform chiral motion due to constant interactions with the obstacles and the diffusivity is reduced until the conditions are right for the motion to be channeled in the lattice. When the strength of dynamic noise is increased, the diffusivity is enhanced for all but high densities, which we attribute to the randomization of the chiral orbits. Moreover, the ability of the swimmer to better explore space becomes less dependent on the obstacles. At high densities, dynamic noise suppresses guided transport through channels. Quenched noise introduces disorder in the spatial arrangement of the obstacles, resulting in erratic motion, preempting fixed swimming orbits, and perturbing guided motion through channels.
\begin{figure*}[ht!]
     \centering
     \includegraphics[width=18.7 cm]{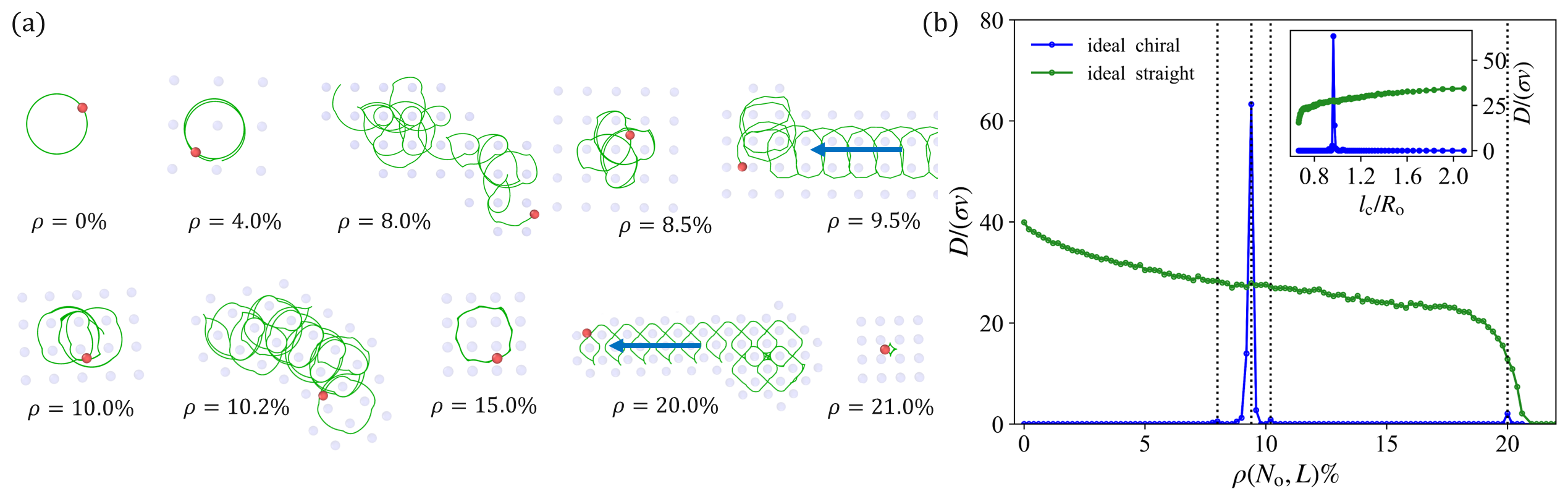}
        \caption{(a) Example trajectories (green) of the swimmer (red) navigating the square lattice of obstacles (faint blue) for increasing densities. The blue arrows indicate the direction of propagation of the `translating orbits'. (b) Diffusivity $D$ vs. density of obstacles $\rho$ for the `ideal chiral swimmer' and the `ideal straight swimmer', which has been rescaled such that for $\rho=0$, $D=40.0$. The inset displays the diffusivity vs. the lattice periodicity $l_{\textrm{c}}$ normalized by the orbit radius $R_{\textrm{o}}$. The dashed lines mark the densities for which diffusion occurs for the `ideal chiral swimmer'.}
        \label{fig_ideal}
\end{figure*}
\section{Model}
We consider a spherically shaped chiral swimmer of diameter $\sigma$ (for \textit{E. coli} typically 1-2 $\mu$m) on a surface containing $N_{\textrm{o}}$ fixed obstacles also of diameter $\sigma$. The amount of obstacles is expressed as the percentage of the surface that is covered by the obstacles: $\rho = \frac {\mathrm{N_{\textrm{o}}} \pi \sigma^2}{L^2}$ ($\times 100 \%$), with $L$ being the edge size of the square simulation box. Interactions between the swimmer and the obstacles are defined by the Weeks-Chandler-Anderson potential obtained by truncating a Lennard-Jones potential at its minimum:
\begin{equation}
    V(r_{i}) =
    \begin{cases}
      4 \left[ \left( \frac{\sigma}{r_{i}} \right)^{6} - \left( \frac{\sigma}{r_{i}} \right)^{12} \right] + 1 & \text{for: $r_{i} < 2^{\frac{1}{6}} \sigma$},\\
      0 & \text{for: $r_{i} \geq 2^{\frac{1}{6}} \sigma$},\\
    \end{cases}       
\end{equation}
where $r_{i}$ is the distance between swimmer and obstacle $i$. 
The initial position of the swimmer is randomly chosen in the box while guaranteeing that it does not overlap with any obstacle. The trajectory of the swimmer is then obtained by integrating the following equations:
\begin{equation}\label{eqmotion}
    \begin{dcases}
    & \frac{{dx}}{d t} = v \cos{(\phi)} + F_x, \\
    & \frac{dy}{d t} = v \sin{(\phi)} + F_y, \\
    & \frac{d\phi}{d t} = \omega  + \sqrt{2D_{\textrm{R}}} \: \xi_{\phi},
    \end{dcases}
\end{equation}
where $x,y$ denote the position of the swimmer, $v$ and $\phi$ are its swimming speed and direction, and $F_x$ and $F_y$ are the $x$ and $y$ components of the force $\bf F = - \nabla$ $V$. The angular velocity $\omega$ results in periodic orbiting motion of the swimmer that is counterclockwise for $\omega > 0$ and $D_{\textrm{R}}$ defines the rotational diffusion constant. The stochastic term $\xi_{\phi}$ represents independent white noise with zero mean and unitary variance. The model reproduced well the experimental data for \textit{E. coli} in \cite{forward}. The time evolution of the position of the swimmer is obtained with a second order Runge-Kutta scheme, to guarantee stability at high obstacle densities. 
As a reference swimmer, we consider an `ideal chiral swimmer' with $v=3\sigma$ and $\omega=1$, representing a swimmer that moves in perfect circles of radius $R_{\textrm{o}} = v/\omega = 3\sigma$ when moving in a homogeneous environment (no obstacles and $D_{\textrm{R}}=0$) as in Fig. \ref{fig_1} (a). As reference surface topography, we consider obstacles placed in a square lattice arrangement (Fig. \ref{fig_1} b). 

Two types of noise are then introduced independently: dynamic noise in the dynamics of the swimmer ($D_{\textrm{R}} \neq 0$) (Fig. \ref{fig_1} c) and quenched noise either in the positions of the obstacles (Fig. \ref{fig_1} d) or in the form of obstacle size polydispersity (Fig. \ref{fig_1} e). For positional quenched noise, the strength is set by $\xi_{\textrm{q}}$, which defines the radius of a circular disk centered at the position of the obstacle; the perturbed obstacles are placed at a new random position chosen uniformly in this disk. For the obstacle size polydispersity, the size distribution follows a Gaussian with mean $\sigma$, and dispersion $\sigma_{\textrm{s}}$. For each obstacle, the diameter $d$ is drawn from the size distribution, if $d \leq 0$ a new diameter is drawn from the distribution until a diameter $d > 0$ is obtained.

In the following we will express distances in terms of the dimensionless obstacle radius $\sigma/2$ and time in terms of $\sigma/v$. A simulation typically includes $N_{\textrm{o}} = 400$ obstacles. Finally, the step size in the simulation is $\Delta t = 10^{-4}$ and a simulation lasts for $t$ = 4800 $\sigma/v$.
\begin{figure}[ht!]
\centering
  \includegraphics[height=17.9cm]{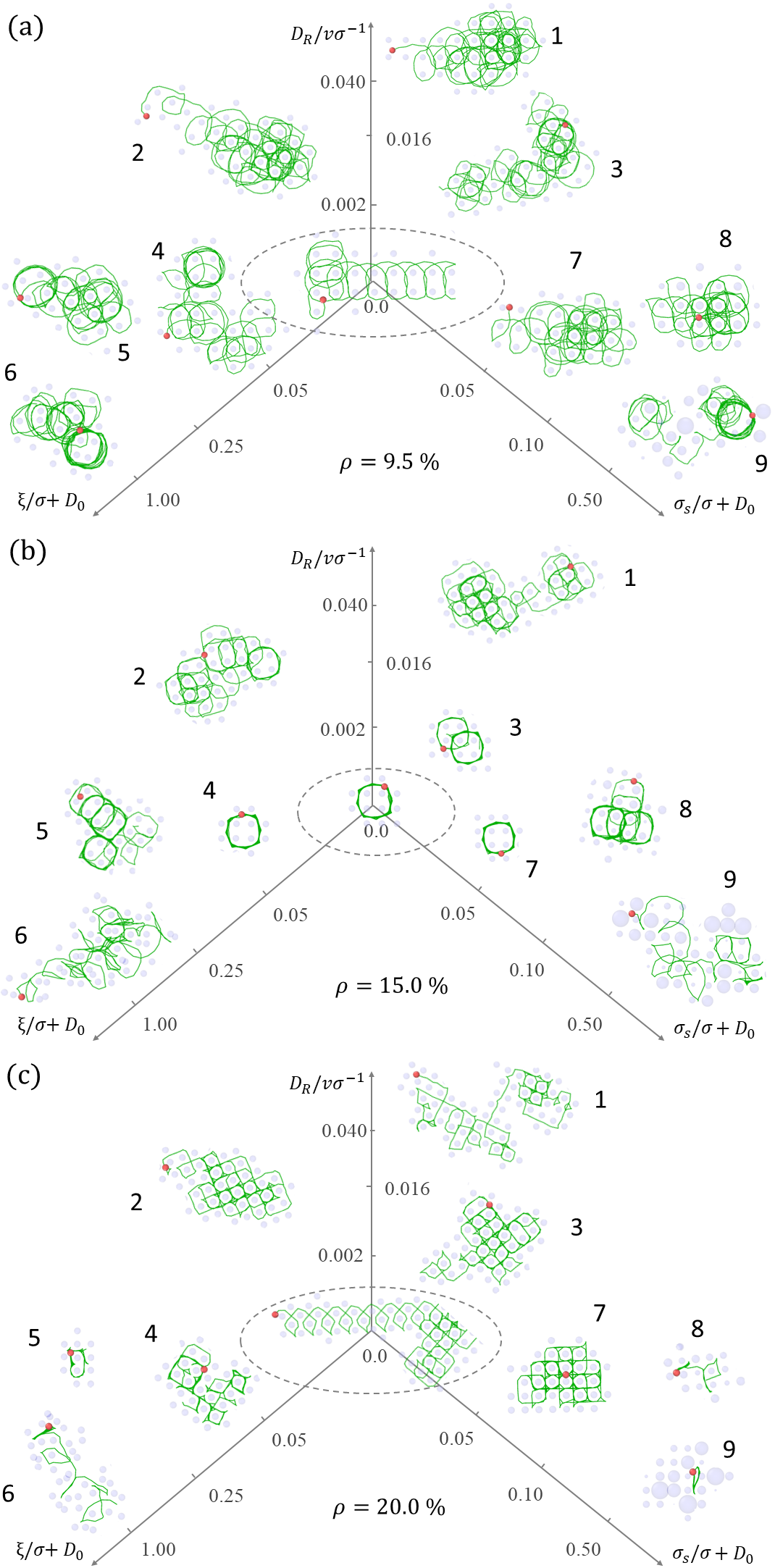}
  \caption{Example trajectories as a function of the different types of noise, for three increasing values of obstacle density: $(a)$ $\rho = 9.5 \%$, $(b)$ $\rho = 15.0 \%$ and $(c)$ $\rho = 20.0 \%$. The swimmer (red) traces a trajectory (green) moving among the obstacles (faint blue). In the middle, in the ellipsoid shape, the trajectory for the `ideal chiral swimmer' ($D_{\textrm{R}}=0$) in a square lattice, with no quenched noise, is shown as reference. The axes indicate the three different noise types: the vertical axis indicates the dynamic noise ($D_{\textrm{R}}$), the axis to the left the positional quenched noise ($\xi$) for a swimmer with $D_{\textrm{0}} \equiv D_{\textrm{R}}/v\sigma^{-1} =0.002$, and the axis to the right quenched noise in the form of size polydispersity ($\sigma_{\textrm{s}}$).}
  \label{fig_trajectories}
\end{figure}

\section{Results}
To characterize the transport properties of the micro-swimmer, we computed the mean square displacement:
\begin{equation}
    \text{MSD}(t) = 
\left\langle [R(t) - R(0) ]^2 \right\rangle,
\end{equation}
with $R(t)$ denoting the position of the swimmer at time $t$. The brackets $\langle . \rangle$ indicate an ensemble average over 4000 swimmer trajectories, each beginning at a position selected uniformly at random. For positional quenched noise or size polydispersity, the average is performed over 100 obstacle configurations, with 40 swimmer trajectories per configuration. 
At long times, the mean squared displacement is expected to scale with the effective diffusion constant $D$ or diffusivity,
\begin{equation}
   \text{MSD} (t) \sim 4Dt,  \quad t \rightarrow \infty.
\end{equation}
which we extracted through a linear regression in the interval $3600\sigma/v < t < 4800\sigma/v$.
The diffusivity is measured for obstacle densities ranging from $\rho=0 \%$ representing a homogeneous environment (Fig. \ref{fig_ideal} (a) for $\rho=0 \%$) to $\rho=21.0\%$, where the swimmer rapidly gets stuck in between obstacles (caged) and is unable to explore space (Fig. \ref{fig_ideal} (a) for $\rho = 21.0 \%$).
\subsection{Ideal chiral swimmer} 
In the case of an `ideal chiral swimmer' ($D_{\textrm{R}} = 0$) in a homogeneous environment (without obstacles), the swimmer moves in a circular orbit of radius $R_{\textrm{o}} =3\sigma$. More interesting behaviour emerges for a heterogeneous environment. A defining feature of the `ideal chiral swimmer' is that its space exploration is deterministic and a result of being scattered by the obstacles. The spatial distribution of obstacles alone governs its dynamics. 

When an `ideal chiral swimmer' navigates a square lattice, the swimmer performs periodic orbits in between or around obstacles at most obstacle densities and is limited to exploring a small part of the system (or `localized'), meaning that it does not perform long time diffusion. Example trajectories of `localized' swimmers are presented in Fig. \ref{fig_ideal} (a) for $\rho = 0\%$, $\rho = 4.0\%$, $\rho = 8.5\%$, $\rho = 10.0\%$, $\rho = 15.0\%$ and a caged swimmer for $\rho = 21.0\%$. For a small number of intermediate densities the swimmer explores space, performing long time `diffusive' behaviour. Examples of diffusive behaviour are shown in Fig. \ref{fig_ideal} (a) for $\rho \approx 8.0\%$, $\rho \approx 9.5 \%$, $\rho \approx 10.2\%$ and $20 \%$. In Fig. \ref{fig_ideal} (b), the diffusivity $D$ of the `ideal chiral swimmer' vs. $\rho $, is presented.   

The sharp peak in the diffusivity at $\rho = 9.5 \%$ is a consequence of space exploration in the form of regular periodic motion. For densities around $\rho = 9.5\%$, the distance in between neighbouring obstacles (the lattice periodicity, Fig. \ref{fig_1} b) $l_{\textrm{c}}$ approximates the radius of the swimming orbit $R_{\textrm{o}}$ causing the swimmer to collide frequently with the surrounding obstacles. Around density $\rho = 9.5\%$, where $R_{\textrm{o}} \approx l_{\textrm{c}}$ (Fig. \ref{fig_ideal} b), the swimmer collides with an obstacle before it completes an orbit, and it is scattered forward towards the next row of obstacles. This `translating orbit' (a term we borrow from \citet{skyrmion}, where a similar effect was observed for skyrmions) is repeated for each row of obstacles, and results in the swimmer efficiently traversing the system. In Fig. \ref{fig_ideal} (a), an example trajectory is presented for $\rho = 9.5\%$. The blue arrow indicates the direction of propagation of the `translating orbit'. At the two much smaller neighbouring peaks, for densities $\rho = 8.0\%$ (Fig. \ref{fig_ideal} a) and $\rho = 10.2\%$ (Fig. \ref{fig_ideal} a), the swimmer is scattered by the obstacles and traces mostly erratic trajectories with occasional laps of periodic behaviour.  For higher densities, at $\rho = 20.0\%$, the swimmer is unable to trace chiral trajectories due to constant interactions with the closely spaced obstacles. Instead, it is guided through channels in the lattice, a trajectory traces a `translating orbit' as the `ideal chiral swimmer' moves through the lattice (Fig. \ref{fig_ideal} a). Efficient transport in the form of `translating orbits', resulting from a crowded topography, strongly contrasts with the behaviour of `ideal straight swimmers' (swimmers with $D_{\textrm{R}}=0$ and $\omega = 0$), for which increasing the obstacle density monotonically hinders space exploration (Fig. \ref{fig_ideal} b).

\subsection{Noise in the dynamics} 
When a chiral swimmer exhibits noisy dynamics ($D_{\textrm{R}} \neq 0$), the circular trajectory is perturbed allowing the swimmer to explore space. For a swimmer experiencing rotational diffusion with strength $D_{\textrm{R}}$, $\tau_{\textrm{R}} = D_{\textrm{R}}^{-1}$ sets the typical time for the swimmer to decorrelate from its initial direction of motion in free space.

Let us now consider a chiral swimmer moving through a square lattice of obstacles. In Fig. \ref{fig_trajectories} (a-c) some example swimmer trajectories are presented for different densities of obstacles. The noise in the dynamics is represented by the vertical axis ($D_{\textrm{R}}/v\sigma^{-1}$). In Fig. \ref{fig_3} (a) the diffusivity is presented for different noise amplitudes $D_{\textrm{R}}$. The dashed line indicates the diffusivity for the swimmer in a homogeneous environment (without obstacles); by comparing the diffusivity for a given noise strength to the homogeneous case $D(D_{\textrm{R}},\rho=0)$, we can determine the effect of the obstacles on the diffusivity.  When a diffusivity curve dips below the dashed line, the presence of obstacles hinders diffusion.

For the smallest noise strength ($D_{\textrm{0}} \equiv D_{\textrm{R}}/v\sigma^{-1} =0.002$), the diffusivity profile is clearly non-monotonic with increasing obstacle density (Fig. \ref{fig_3} a). The noise perturbs the localized orbits of the `ideal chiral swimmer', resulting in diffusive behaviour for all densities $\rho < 20.5 \%$. For higher densities, the swimmer is caged by the obstacles and the diffusivity vanishes.  A fine structure with four peaks emerges, around densities $\rho = 2.0 \%$, $\rho = 4.5 \%$, $\rho = 9.5 \%$ and $\rho = 20.0 \%$. Less pronounced minima occur at $ \rho = 2.5 \% $, $\rho = 6.0 \%$ and $ \rho = 15.0\%$. For densities $\rho < 7.5\%$, the fine structure does not qualitatively change the behaviour of the swimmer, only resulting in a small enhancement or suppression of the diffusion. The fine structure is a result of the chiral swimmer interacting with the periodic lattice, and disappears if the lattice is sufficiently perturbed, which will be explored in depth in paragraph \ref{posnoise}.

To understand the main features of the diffusivity (Fig. \ref{fig_3} a), we can look at the number of orbits $n = \frac{\omega}{2\pi D_{\textrm{R}}}$ a swimmer typically completes (in a homogeneous environment) before its orientation is randomized by the dynamic noise, as suggested by \cite{franosh2}. For the small noise strength ($D_{\textrm{R}}/v\sigma^{-1}=0.002$), we obtain $n =159$, implying that the swimmer still largely traces a circular trajectory if it does not interact with the obstacles. In this case, the increased diffusion can be attributed to efficient scattering by obstacles, rectifying the swimming orbits, and enhancing the swimmer's diffusion. This mechanism dominates for intermediate densities, especially around $\rho = 9.5\%$ where the radius of the swimming orbit $R_{\textrm{o}}$ approximates the spacing between obstacles $l_{\textrm{c}}$, such that $l_{\textrm{c}}/R_{\textrm{o}} \approx 1$ (Fig. \ref{fig_3} a inset). Some example trajectories are presented along the vertical axis ($D_{\textrm{R}}/v\sigma^{-1}$) in Fig. \ref{fig_trajectories} (a.1-3). The suppression of the diffusion for densities around $\rho = 15.0\%$ results from the swimmer becoming intermediately trapped in trajectories that orbit around a few obstacles. Due to the interactions with the obstacles, the swimmer keeps moving in a fixed orbit until, driven by dynamic noise, it hops (diffuses) to the next row of obstacles. Similar orbits were also observed for the `ideal chiral swimmer' for densities around  $\rho = 15.0 \%$, but since the `ideal chiral swimmer' does not experience dynamic noise, it remains trapped indefinitely (Fig. \ref{fig_ideal} for $\rho = 15 \%$). In Fig \ref{fig_trajectories} (b.3), an example trajectory is presented of a swimmer that hopped to a new orbit once. At higher densities the channels narrow further; now a small deviation in the trajectory more readily results in the swimmer hopping across rows, enhancing diffusion.
Around density $\rho = 20.0 \%$ (the highest densities before caging occurs), the frequent interactions with the closely spaced obstacles prevent the swimmer from moving in circular orbits. Instead, the swimmer is guided through channels in the lattice, in an erratic way, efficiently exploring space. In Fig. \ref{fig_trajectories} (c.1-3) some example trajectories of a swimmer with dynamic noise navigating a dense system with $\rho = 20.0 \%$ are presented. For higher densities $\rho \geq 20.5 \%$, the swimmer is fully caged in the lattice.

Now that the general behaviour of a noisy chiral swimmer is outlined, we can examine the effect of increasing the noise strength $D_{\textrm{R}}$ (Fig. \ref{fig_3} a). As a reference, we take the previous case of a swimmer with $D_{\textrm{R}}=D_{\textrm{0}}$. For a small increase in the noise strength to $D_{\textrm{R}}/v\sigma^{-1}=0.004$, and $n=80$, the behaviour is similar to the reference case. When the noise is increased to $D_{\textrm{R}}/v\sigma^{-1}=0.016$, we obtain $n=20$; now the presence of obstacles enhances the diffusivity for all densities in a more uniform manner, but still peaks around $\rho = 9.5 \%$, before flattening out. Limited enhancement is observed for densities  $\rho > 15 \%$ until it vanishes for $\rho \geq 20.5 \%$. In this case, the noise is too strong for the swimmer to remain trapped in fixed periodic orbits for an extended period of time. Instead, the swimmer moves across the lattice in an erratic way, along with intermittent periods of orbiting around (multiple) obstacles (Fig. \ref{fig_trajectories} (a-c.2)). The diffusion of such erratic motion is not very sensitive to the density; the diffusivity varies little for $\rho > 15.0 \%$, until it vanishes at $\rho = 20.5 \%$ as the swimmer becomes caged by the obstacles. 

For still larger noise strengths ($D_{\textrm{R}}/v\sigma^{-1}=0.040$ and $D_{\textrm{R}}/v\sigma^{-1}=0.080$), the noise increasingly perturbs the orbiting motion, the swimmer is markedly less chiral, typically only able to perform $n=8$ orbits ($D_{\textrm{R}}/v\sigma^{-1}=0.040$) and $n=4$ orbits ($D_{\textrm{R}}/v\sigma^{-1}=0.080$) before its orientation is fully decorrelated. In this case, the swimmer traces strongly erratic trajectories for all densities, which are not very sensitive to the density. Some example trajectories are presented in Fig. \ref{fig_trajectories} (a-c.1) for $D_{\textrm{R}}/v\sigma^{-1}=0.040$. Scattering by obstacles enhances diffusion only marginally for low densities $\rho < 9.5 \%$ ($D_{\textrm{R}}/v\sigma^{-1}=0.040$) and $\rho < 3.0 \%$ ($D_{\textrm{R}}/v\sigma^{-1}=0.080$).  Moreover, the diffusivity varies little with the density for $\rho < 9.5 \%$ ($D_{\textrm{R}}/v\sigma^{-1}=0.040$) and $\rho < 3.0 \%$ ($D_{\textrm{R}}/v\sigma^{-1}=0.080$). For higher densities, similarly to the `straight swimmer' (black curve in Fig. \ref{fig_3} a), a monotonic decrease in diffusivity for increasing density is retrieved, vanishing for $\rho \geq 20.5 \%$ where the swimmer is caged in the lattice.

\subsection{Randomness in the positions of the obstacles}\label{posnoise}
When a chiral swimmer with $D_{\textrm{0}}$ explores a lattice with noise in the positions of the obstacles, the diffusivity is markedly affected for large strengths of noise. In Fig. \ref{fig_3} (b), the diffusivity is presented for different amplitudes of the noise $\xi_{\textrm{q}}$, alongside the diffusivity of a configuration with randomly placed non-overlapping obstacles. The largest noise strength ($\xi_{\textrm{q}}/\sigma = 1.00$) corresponds to configurations where the contact between obstacles at density $\rho=20.0 \%$ is still low, with typically less than $10.0 \%$ of the obstacles touching. In Fig. \ref{fig_trajectories}, noise in the positions is represented by the left axis ($\xi/\sigma + D_{\textrm{0}}$).

For the smallest noise strength ($\xi_{\textrm{q}}/\sigma=0.05$), at low densities ($\rho < 7.0 \% $), the fine structure is little affected. The diffusivity profile is affected for high densities ($\rho > 17.0 \% $), with the peak around $\rho = 20.0 \%$ being significantly suppressed and shifted towards lower density values with respect to the ordered lattice ($\xi_{\textrm{q}}/\sigma=0)$. The shift in the position of the peak around $\rho = 20.0 \%$ is a result of the obstacles guiding the swimmer through (disordered) channels, leading to a small local maximum in the diffusivity for densities around $\rho = 19.5 \%$. For higher densities, the randomness combined with the densely spaced obstacles perturbs the channels, and introduces significant backtracking of the trajectories, reducing the diffusivity when compared to the ordered lattice (Fig. \ref{fig_trajectories} (c.4)). At the highest densities $\rho > 19.5 \%$, the obstacles are closely spaced; even a small perturbation of the channels introduces bottlenecks a swimmer can not pass through, limiting diffusion.

When the noise is increased to $\xi_{\textrm{q}}/\sigma = 0.25$, the peak is shifted towards lower densities, channels are randomized further and bottlenecks occur at lower densities. Now, the disordered obstacles preempt the swimmer from getting trapped for long in intermittently stable orbits around a number of obstacles, as observed in the ordered lattice for densities around $\rho = 15.0 \%$ (Fig. \ref{fig_trajectories} (b.4) and for the `ideal chiral swimmer' \ref{fig_ideal} a). The swimmer is scattered randomly and diffuses more efficiently (Fig. \ref{fig_trajectories} (b.5)), reducing the depth of the local minimum in the diffusivity. As the lattice perturbation is increased, bottlenecks occur at lower densities ($\rho > 18.0 \%$). An example trajectory is presented in Fig. \ref{fig_trajectories} (c.5). 

For large amplitudes of positional noise ($\xi_{\textrm{q}}/\sigma = 0.50$ and $\xi_{\textrm{q}}/\sigma = 1.00$), with the average random displacement comparable to the obstacle size, the lattice is strongly perturbed. This results in a smoothing out of the fine structure of the diffusivity profile for all densities. In the limit of large positional quenched noise, a random topography is retrieved, and the diffusivity exhibits a single flattened peak, centered around $\rho = 10.0 \%$ (black curve in Fig. \ref{fig_3} b). Due to the positional disorder, the distance between obstacles becomes irregular. This leads to suppression of the diffusion for densities around $\rho = 9.5\%$, as the disordered arrangement preempts regular motion where the swimmer moves from one row of obstacles to another, instead it is scattered randomly. As the lattice is further perturbed, this effect becomes increasingly pronounced (Fig. \ref{fig_trajectories} (a.4-6)). Even for a fully random arrangement, the diffusion is enhanced for  densities $\rho < 17.0 \%$, indicating that for all but high densities of randomly placed obstacles, the rectification of swimming orbits by obstacles leads to enhanced diffusion. For large amplitudes of positional noise (and the random configuration), and increasing densities $\rho > 17.0 \%$, the strongly disordered and densely packed arrangements increasingly contain pockets of obstacles that trap the swimmer, limiting diffusion. A trajectory of a swimmer in a strongly disordered lattice that ultimately becomes trapped in a fixed orbit is displayed in Fig. \ref{fig_trajectories} (c.6).

\subsection{Non-uniform size distribution}
\begin{figure}[ht!]
\centering
  \includegraphics[height=17.2cm]{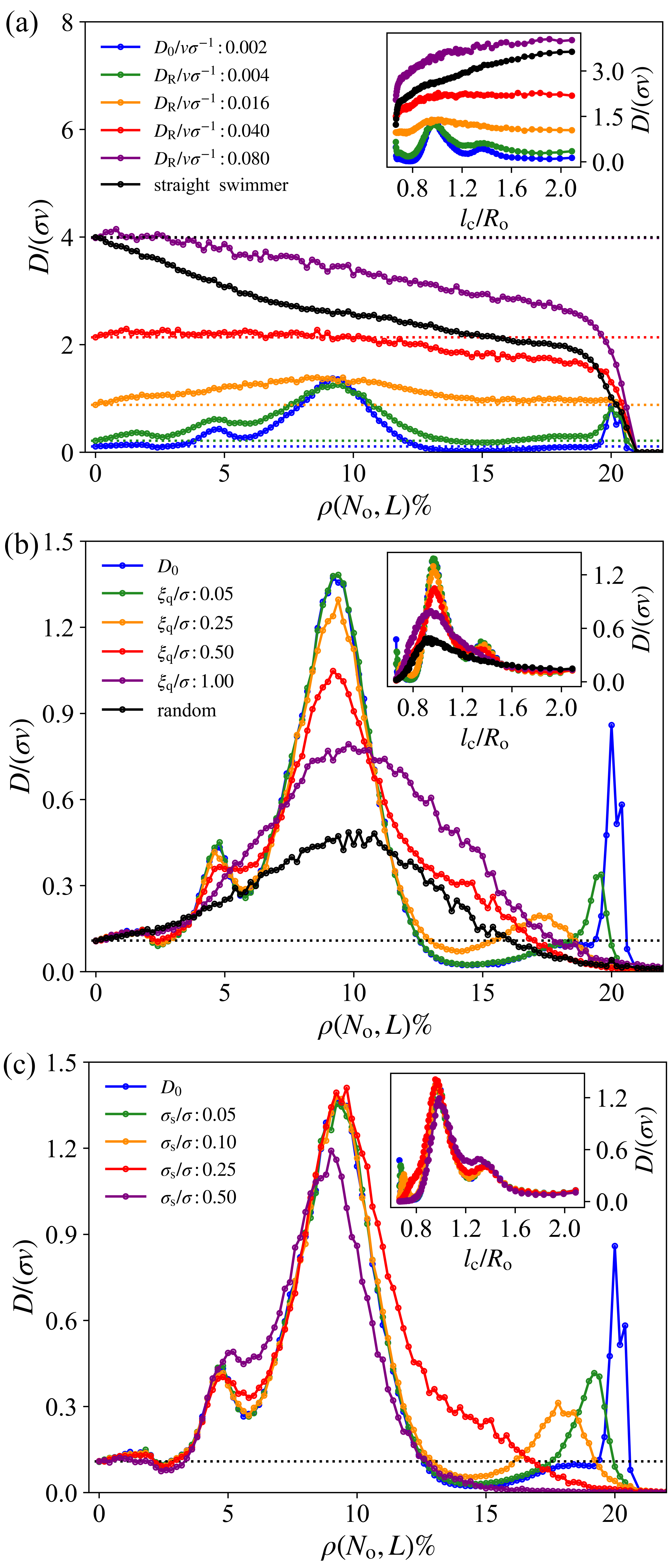}
  \caption{Diffusivity $D$ vs. density of obstacles: (a) for different dynamic noise amplitudes $D_{\textrm{R}}$, (b) with $D_{\textrm{R}}/v\sigma^{-1}=D_{\textrm{0}}=0.002$ for different lattice perturbation amplitudes $\xi_{\textrm{q}}$, (c) with $D_{\textrm{R}}/v\sigma^{-1}=D_{\textrm{0}}=0.002$ for different amplitudes of size polydispersity $\sigma_{\textrm{s}}$. The inset displays the diffusivity vs. the lattice periodicity $l_{\textrm{c}}$ for the obstacles, for a non perturbed lattice, normalized by the orbit radius $R_{\textrm{o}}$. The dashed lines indicate the diffusivity for a system without obstacles, and serve as a guide for the eye. The straight swimmer in (a) represents a swimmer with with velocity $v=3\sigma/s$, $\omega =0$ and $D_{\textrm{0}}$, and has been rescaled such that, for $\rho=0$, it matches the curve with $D_{\textrm{R}}/v\sigma^{-1}=0.080$. The random system in (b) represents a regular chiral swimmer with $D_{\textrm{0}}$, in a system with randomly placed non overlapping obstacles.}
  \label{fig_3}
\end{figure}

For a non-uniform obstacle size distribution, $\sigma_{\textrm{s}}$ sets the strength of the obstacle size polydispersity. In Fig. \ref{fig_3} (c), the diffusivity is presented for different $\sigma_{\textrm{s}}$. Here the limit of large polydispersity ($\sigma_{\textrm{s}}/\sigma = 0.50$) corresponds to configurations with about $20.0 \%$ of obstacles, typically the largest, touching at density $\rho = 20.0 \%$.  The right axis in Fig. \ref{fig_trajectories} represents the size polydispersity ($\sigma_{\textrm{s}}/\sigma + D_{\textrm{0}}$).

When navigating a surface with polydisperse obstacles, the diffusivity profile for a swimmer with $D_{\textrm{0}}$ is affected similarly to the system with positional quenched noise. For the smallest polydispersity ($\sigma_{\textrm{s}}/\sigma = 0.05$), at densities $\rho < 17.0 \% $, the fine structure is little affected, as the swimmer traverses the system in an erratic way (Fig. \ref{fig_trajectories}(a.7) provides an example for $\rho = 9.5 \%$). For higher densities $\rho > 17.0 \% $, similar to the positional noise, the shift in the position of the peak around $\rho = 20.0 \%$ is a result of the obstacles guiding the swimmer through channels of irregularly sized obstacles (similar to Fig. \ref{fig_trajectories} (c.7)). This leads to a local maximum in the diffusivity for densities around $\rho = 19.0\%$. For still higher densities bottlenecks and pockets form that trap the swimmer. When the polydispersity is increased slightly to $\sigma_{\textrm{s}}/\sigma = 0.10$, the effect of intermittent trapping in orbits for densities around $\rho = 15.0 \%$ is reduced (Fig. \ref{fig_trajectories} (b.8)) and the peak is smeared out and shifted to lower densities. 

For polydispersity  $\sigma_{\textrm{s}}/\sigma = 0.25$, the fine structure is smoothed out for $\rho > 12.0 \% $ as the swimmer explores the system efficiently tracing erratic trajectories, until it increasingly becomes trapped for $\rho > 18.0 \%$ and the diffusivity vanishes (Fig. \ref{fig_trajectories} (c.8)). When the polydispersity is increased to $\sigma_{\textrm{s}}/\sigma = 0.50$, the fine structure is smoothed out for all densities, the peak around $\rho = 9.5 \%$ is shifted to slightly lower densities and is reduced as the trajectories are randomized and periods of intermittent trapping in orbits occur (Fig. \ref{fig_trajectories}(a.9)). The shifting of the peak can be explained by the larger total surface covered by the obstacles due to the size polydispersity (what is gained by the obstacles that increase in size is not offset by what is lost by the obstacles that become smaller). For $\sigma_{\textrm{s}}/\sigma = 0.50$ the diffusivity is suppressed relatively to $\sigma_{\textrm{s}}/\sigma = 0.25$, for almost all densities. This is a result of some of the obstacles becoming increasingly larger; for $\rho > 8.0 \%$ the large obstacles form clusters that trap the swimmer, as well as walls that subdivide the system and prevent the swimmer from accessing parts of it. For $\rho > 15.0 \%$, the diffusivity vanishes rapidly as the swimmer becomes caged in pockets of obstacles. In Fig. \ref{fig_trajectories}(a-c.9), some example trajectories are displayed for noise amplitudes $\sigma_{\textrm{s}}/\sigma = 0.50$. When the polydispersity is further increased ($\sigma_{\textrm{s}}/\sigma > 0.50$), some obstacles become much larger than the swimmer; in this limit swimmer-obstacle interactions become more important, and our simplified model is no longer expected to accurately capture the dynamics \cite{following1, following2}.

\section{Conclusions}
We have investigated the dynamics of a chiral micro-swimmer on a surface with obstacles. Different densities of obstacles as a fraction of the surface area were considered. We introduced distinct types of disorder: noise in the dynamics of the swimmer and noise in the positions of the obstacles, as well as obstacle size polydispersity. 

A noiseless chiral swimmer in an environment without obstacles gets trapped in circular orbits and does not explore space. When navigating an array of regularly spaced obstacles, for most densities, space exploration of a chiral swimmer is limited to fixed swimming orbits. Efficient long time transport, in the form of a `translating orbit', occurs when the distance between the obstacles approximates the orbit radius of the swimmer, or the spacing between the obstacles approximates the size of the swimmer. Less efficient long time transport occurs for a few densities, with the swimmer tracing an erratic trajectory.

A chiral swimmer that experiences dynamic noise performs long time transport for all densities, except for very high densities when it becomes trapped by the obstacles. The presence of low to intermediate densities of obstacles enhances the transport of a chiral swimmer, by rectifying swimming orbits. Especially when the distance between the obstacles approximates the radius of the swimmer orbit, this effect is pronounced. For higher densities, obstacles can enhance transport by guiding the swimmer through open channels in the lattice. 

When the strength of the dynamic noise is increased, the chiral trajectories are more strongly perturbed. This reduces the tendency of the swimmer to swim in circles, enhancing transport for all densities, and reducing the effect of scattering by obstacles on the transport. For large noise amplitudes, the swimmer is effectively no longer chiral. In this case, the presence of obstacles hinders transport, which is in agreement with previously published results for non-chiral swimmers \cite{experiments1, lorentzgas}. 

Positional noise and obstacle size polydispersity result in a disordered environment, which enhances transport by preempting fixed swimming orbits in between or around obstacles. Additionally, scattering of the swimmer by the disordered obstacles results in erratic motion. For densities where the inter-obstacle distance approximates the radius of the swimmer orbit, such motion perturbs the otherwise efficient rectification of swimming orbits, and suppresses transport. 

For high densities, quenched noise perturbs channels, limiting, and suppressing transport. For high densities and large quenched positional disorder or size polydispersity, pockets of obstacles form that trap the swimmer, strongly hindering transport. In the limit of large positional noise, the presence of a low or intermediate density of obstacles enhances transport. When the obstacle size polydispersity is increased, the dynamics is increasingly hindered by a few very large obstacles that form pockets that trap the swimmer, inhibiting long term transport. Moreover, when the obstacles are much larger than the swimmers, often the swimmers tend to move along the boundary of the obstacles, and in this limit our model might no longer be accurate \cite{following1,following2}. To describe the dynamics of such systems, a model that explicitly incorporates a boundary following interaction, where a swimmer is guided along the boundary of an obstacle upon contact, such as in \cite{franosh1, franosh2}, should be more realistic.

An interesting extension of this work would be to investigate the effect of spatial arrangement and disorder on the dynamics of interacting chiral micro-swimmers. Previous work encountered motility induced phase separation, commensuration effects and frustrated states for a simple model of interacting non-chiral swimmers in a similar square lattice of obstacles \cite{frustration}. Another possibility would be to consider dynamic obstacles. The bronchus in the lungs are lined with respiratory cilia: microscopic, periodically beating hairs that clear inhaled debris and microbes from the conducting airways \cite{cilia}. By introducing a periodic motion in the obstacles, the dynamics governing the transport of (chiral) microbes in the cilia could be explored.  Although this work is limited to surface topography, extending it to three dimensions would also be interesting. Recently a geometric criterion for the optimal transport of run and tumble polymers in a porous medium was uncovered, which emerges when their run lengths are comparable to the longest straight path available in the porous medium \cite{3d}. From a very different perspective, simulation studies of circular ac driven skyrmions on a square lattice of obstacles encountered behaviours similar to our findings for the `ideal circle swimmer'\cite{skyrmion}. It could be interesting to see how such systems respond to noise. In addition to a fundamental understanding of bacterial dynamics near surfaces and interfaces as well as of search strategies and optimal navigation in complex environments, this work can be employed to guide the design microstructured surfaces that can control and prevent bacterial adhesion. 

\section*{Acknowledgements}
We acknowledge financial support by the European Commissions Horizon 2020 research and innovation program under the Marie Sklodowska-Curie Grant Agreement No. 812780 and from the Portuguese Foundation for Science
and Technology (FCT) under Contracts no. PTDC/FIS-MAC/28146/2017 (LISBOA-
01-0145-FEDER-028146), UIDB/00618/2020, and UIDP/00618/2020.


\bibliography{main} 
\bibliographystyle{main} 

\end{document}